\documentclass{article}
\usepackage{arxiv}
\pdfoutput=1 
\usepackage[utf8]{inputenc} % allow utf-8 input
\usepackage[T1]{fontenc}    % use 8-bit T1 fonts
\usepackage{hyperref}       % hyperlinks
\usepackage{url}            % simple URL typesetting
\usepackage{booktabs}       % professional-quality tables
\usepackage{amsfonts}       % blackboard math symbols
\usepackage{dsfont} % doublestroke
\usepackage{microtype}      % microtypography
\usepackage{lipsum}         % Can be removed after putting your text content
\usepackage{doi}
\usepackage{graphicx}
\usepackage{blkarray}
\usepackage{amsmath,amssymb}
\usepackage{pdflscape}
\usepackage{longtable}
\usepackage{multirow}
\usepackage{xcolor}
\usepackage{calc}
\usepackage{booktabs}
\usepackage{colortbl}
\usepackage{mathtools}
\usepackage[authoryear,round,sort,comma,numbers]{natbib}
\usepackage{helvet}
\usepackage[shortlabels]{enumitem}

\title{Multiple State Analysis, a Multidimentional Approach to Multiple Time-to-Event Data and\\ Life Course Health Trajectories:\\  Application to Patients with Myocardial Infarction}
\author{
\href{https://orcid.org/0000-0002-0455-6749}
{\includegraphics[scale=0.06]{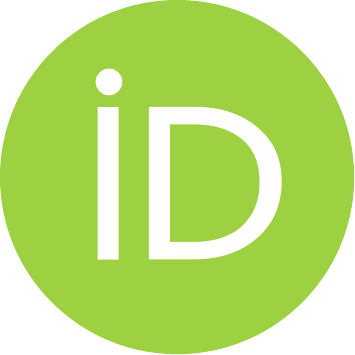}\hspace{1mm}Marc Delord}
\thanks{Correspondence: marc.delord@kcl.ac.uk}\\
School of Life Course \& Population Sciences\\
Department of Population Health Sciences\\ King's College London\\ London, United kingdom\\
\And
\href{https://orcid.org/0000-0001-6964-9041}
{\includegraphics[scale=0.06]{orcid.pdf}\hspace{1mm}Annastazia Learoyd}\\
School of Life Course \& Population Sciences\\
Department of Population Health Sciences\\ King's College London\\ London, United kingdom\\
\And
\href{https://orcid.org/0000-0002-4354-4433}
{\includegraphics[scale=0.06]{orcid.pdf}\hspace{1mm}Abdel Douiri}\\
School of Life Course \& Population Sciences\\
Department of Population Health Sciences\\ King's College London\\ London, United kingdom\\
}  

\hypersetup{
pdftitle={Multiple State Analysis applied to Myocardial Infarction},
pdfsubject={stat.ME, stat.AP},
pdfauthor={Marc Delord},
pdfkeywords={Jaccard Dissimilarity Index, Multimorbidity, Multiple Sequence Analysis, Multiple State Analysis, 
Myocardial Infarction, Ward clustering},
}

\renewcommand{\headeright}{\emph{Preprint}}
\renewcommand{\undertitle}{\emph{Preprint}}
\renewcommand{\shorttitle}{Multiple State Analysis applied to Myocardial Infarction}

\begin{document}
\maketitle

\begin{abstract}
Life course epidemiology of chronic diseases has been dominated so far by the environmental approach. Whether it focuses on early life exposures / events or later lifestyle behaviors, this approach assumes that previous life experiences interact at each stage of life and shape subsequent health trajectories. Inspired by the analysis of social trajectories, focusing on transitions between multiple states, in multiple dimensions of the social experience, we propose here a novel empirical approach to multiple time-to-event health data, denoted as \emph{multiple state analysis}. Alike the so-called state sequence analysis, the aim of \emph{multiple state analysis} is to create typologies of the main life course trajectories. This approach is applied to real world data from a south London general practice electronic health record system from which multiple long term conditions associated with myocardial infarction were considered. Among expected results such as the recurrent role of hypertension, \emph{multiple state analysis} shows that different patterns of long term conditions including physical and mental health conditions, are associated with the onset of myocardial infarction but also with socio-demographics such as sex and ethnicity.
\end{abstract}

% keywords can be removed

\section{Introduction}

The notion that life course health trajectories are influenced by early life events and the ever-changing historical context \citep{Elder1998}, or by later \emph{lifestyle} associated exposure \citep{Hill1957} have dominated the epidemiological field so far. Conversely, we can assume that patterns induced by the main combinations of early life events and specific socio demographic contexts, would produce tractable patterns of the main life-course health trajectories. Progress in information technologies and the availability to academic research of publicly funded health databases including daily life general practice records represent an unprecedented opportunity to better evaluate this hypothesis. Along with the main trajectories which are well documented such as the association of hypertension with cardio-vascular diseases \citep{Carter2015}, this research is expected to better identify critical trajectories and potentially uncover unexpected patterns and at-risk populations allowing better informed healthcare policies.

Whereas life course epidemiology has developed the environmental model-based approach to chronic diseases, focusing  on lifestyles and exposures, the social sciences have proposed a sequential approach of life experience trajectories. This empirical approach focuses on transitions between distinct states and aims at creating typologies of trajectories by identifying patterns of recurrent sequences. In subsequent developments of the so-called state sequence analysis, Pollock \citep{Pollock2007} proposed an extended approach to life course trajectories denoted as \emph{multiple sequence analysis} where multiple dimensions of the social experience are considered when building a typology.

In this paper, we propose a generalization of Pollock's approach, to  multiple time-to-event endpoints through the analysis of multiple long term conditions associated with myocardial infarction. Myocardial Infarction, leading to ``heart attack," is one of the leading causes of death in high-income countries \citep{roth2015global}. Myocardial infarction is caused by decreased or complete cessation of blood flow in the myocardium and results in irreversible damage to the heart muscle \citep{jaffe2013third}. Most of the time, myocardial infarction is due to underlying coronary artery disease \citep{Libby2013}. Well-known modifiable risk factors associated with coronary artery disease and myocardial infarction are smoking, abnormal blood lipid profile, hypertension, diabetes mellitus, abdominal obesity, psycho-social factors, diet, physical activity and alcohol consumption (protective) \citep{Yusuf2004,Anand2008}. Some non-modifiable risk factors for myocardial infarction include advanced age, male gender (males tend to have myocardial infarction earlier in life) and genetics \citep{Anand2008, Nielsen2013}.

This paper is organized as follows: section 1 presents the background of the analysis of state sequences, a motivating example justifying the use of the proposed approach and ends by the presentation of the \emph{multiple state analysis} method. In section 2 the analysis of multiple long term conditions associated with myocardial infarction is presented. Finally in section 3 the \emph{multiple state analysis} method is briefly discussed in light of the study results.

\subsection{ State Sequence Analysis }

State sequence analysis is the analysis of sequences describing individual life courses with the objective to develop exploratory typologies of the main trajectories at the population level. As this empirical approach would always produce some results \citep{Levine2000}, and bearing in mind that state sequence analysis is an exploratory exercise, the sound interpretation of resulting typologies, grounded by relevant theories \citep{Shalizi2009} remains an important step of the state sequence analysis workflow. In the social sciences and epidemiological setting, beside the interest of a typology by itself \citep{Pollock2007}, the use of auxiliary interpretative variables such as socio-demographic covariates or known risk factors, is of particular interest to evaluate the relevance of typologies insofar as public health policies depends on identifiable modifiable risk factors and associated populations. First established in the area of social sciences, state sequence analysis is applicable to domains where temporal or spatial transitions are meaningful, such as in the social sciences, in biology or in healthcare research \citep{Vanasse2020}.

Although state sequence analysis is an empirical method, it is important to state that in the settings mentioned earlier, analysed sequences require some $good$ properties to produce meaningful results. Especially, transitions between states need to be meaningful and successive states should be quasi-exclusive from one another: hospital discharge in healthcare research, disease recovery in epidemiology, or change in marital status in social sciences are examples of meaningful state transitions. In other situations however, the simple record of successive non exclusive events may not fully render subjects' experiences and the analysis of such sequences may be misleading.

For instance, in the medical field, the onset of some communicable or non communicable diseases such as long term conditions, the record of which may not always be fully accurate or the diagnosis based on heterogeneous criteria, will produce such sequences. Importantly, if the sequence of acute health events and long term conditions is informative for single time-to-event analysis, the concept of multimorbidity, that is a state characterized by the accumulation of 2 or more long term conditions, that may occur simultaneously, represents a relevant information in term of life course health trajectories \citep{Barnett2012}. However, this dimension may not be fully captured by the simple sequence of events further analysed by state sequence analysis. 

To illustrate this situation let's consider health record histories of a homogeneous group of patients presenting a diagnosis of diabetes and hypertension at about the age of fifty and a record of stroke at about the age of 60. Though we intuitively understand that these patients share similar experiences in terms of health trajectories, the simple sequences derived from their historical records will present a significant divergence in the critical ten years preceding the onset of stroke, according to whether hypertension or diabetes was first recorded. As a consequence, the simple analysis of these sequences will produce a typology dominated by the nature of the first recorded long term condition  (diabetes or hypertension). This conclusion would be arguably misleading as, by hypothesis, analysed patients show similar health trajectories.

One way to overcome this situation could be to produce some \textit{ad hoc} sequences for these patients such as:
\parbox{\textwidth}{
\begin{enumerate}[(i),itemsep=0mm]
 \item \textit{diabetes} or \textit{hypertension}
 \item $\{$\textit{diabetes \:\: and \:\: hypertension}$\}$ or $\{$ \textit{hypertension \:\: and \:\: diabetes }$\}$
 \item $\{$\textit{diabetes \:\: and \:\: hypertension\ \:\: and \:\: stroke}$\}$ or\\
 $\{$\textit{hypertension \:\: and \:\: diabetes \:\: and \:\: stroke}$\}$,
\end{enumerate}
}
where $\{\:\: . \:\: and \:\: . \:\: and \:\: ... \}$ represent a new states of multimorbidity. However, in this situation, characterized by different states of multimorbidities, the definition of simple sequences may not be straightforward and would require simplifying rules produced by experts of the relevant field in order to obtain a reduced set of workable and meaningful life course health sequences.

\subsection{Multiple State Analysis}\label{MSA}

In a similar situation considering social trajectories Pollock \citep{Pollock2007} proposed the multiple-sequence analysis where sequences associated with distinct dimensions of the social experience are analysed in conjunction with one another. In order to deal with limitations associated with the improper use of state sequence analysis when analyzing non excluding / non competing events, we propose a novel approach related to Pollock's method denoted hereafter as \emph{multiple state analysis} -- or multiple time to event analysis -- applied to the analysis of multimorbidity associated with myocardial infarction as an example.

From a statistical perspective, the proposed method requires a multidimentional representation of patients' health records. In subsequent steps, the classical state sequence analysis principles are applied: given an appropriate metrics, pairwise dissimilarity between patients are computed before a clustering method is applied to create a typology.

\subsubsection{Individual patient's state matrix and group summaries}\label{sm}

\emph{Multiple state analysis} requires individual health patients' records to be formatted into multiple time-to-event indicator tracks, stacked in ${t \times k}$ state matrices, $t$ and $k$ being the maximum age observed in the cohort and the total number of conditions considered in the analysis, respectively.

If onset of disease $l$ for patient $j$ is $t_j$, we note: $\textrm{M}_{.l}^j = \textrm{I}_{[t \geq t_j]}$, where $\textrm{M}_{.l}^j$ stands for column $l$ of matrix $\textrm{M}^j$, and I is the positive integer indicator function. For instance in the following example, $\textrm{M}^j$ shows that the analysis is conducted on long term conditions $a$, $b$, $c$, $d$ and $e$ from 0 to 99 years old for patient $j$, onset of diseases $a$, $b$, $c$, and $e$ being 2, 46, 47 and 48 respectively. Patients may be censored or experience a competing event such as death or any other event whose occurrence either precludes the occurrence of diseases under examination or fundamentally alters the probability of their occurrence \citep{gooley1999estimation}. 

$ \tau $ being patients' censoring times vector, $\textrm{c}^j = \textrm{I}_{[t \geq \tau_j]}$ represents patient's $j$ censoring indicator and $\bar{\textrm{c}}^j = \textrm{I}_{[t < \tau_j]}$, the follow up period for patient $j$. In the proposed example, patient $j$ is censored at age 83.

\begin{equation}\label{mat}
  \mathbf{M^j}=
\begin{blockarray}{cccccc}
& a & b & c & d & e  \\
\begin{block}{c(ccccc)}
0 & 0 & 0 & 0 & 0 & 0 \\
  1 & 0 & 0 & 0 & 0 & 0 \\
  2 & 1 & 0 & 0 & 0 & 0 \\
  \vdots & \vdots & \vdots & \vdots & \vdots & \vdots \\
  45 & 1 & 0 & 0 & 0 & 0 \\
  46 & 1 & 1 & 0 & 0 & 0 \\
  47 & 1 & 1 & 1 & 0 & 0 \\
  48 & 1 & 1 & 1 & 0 & 1 \\
\vdots & \vdots & \vdots & \vdots & \vdots & \vdots \\
81 & 1 & 1 & 1 & 0 & 1 \\
82 & 1 & 1 & 1 & 0 & 1 \\
83 & 1 & 1 & 1 & 0 & 1 \\
\vdots & \vdots & \vdots & \vdots & \vdots & \vdots \\
99 & 1 & 1 & 1 & 0 & 1 \\
\end{block}
\end{blockarray},\:\:
\mathbf{\bar{c}^j} =
\begin{blockarray}{cc}
 &   \\
\begin{block}{c( c )}
0 & 1  \\
1 & 1  \\
2 & 1  \\
\vdots & \vdots \\
45 & 1 \\
46 & 1 \\
47 & 1 \\
48 & 1 \\
\vdots & \vdots  \\
81 & 1 \\
82 & 1 \\
83 & 0 \\
\vdots & \vdots  \\
99 & 0 \\
\end{block}
\end{blockarray} 
\end{equation}

An appealing feature of the proposed individual patients' data representation is its straightforward interpretation in terms of the classical time-to-event indicators such as overall survival and cumulative incidence functions, and associated graphical representations, when interested in statistical summaries associated to the entire cohort or any subgroup of patients/diseases. This is of particular interest in the medical setting where these statistics are routinely interpreted.% Of note, the cause specific hazard ratio of all diseases under investigation is also estimable from this data presentation (supplementary figure 1). 

\subsubsection{Dissimilarity index matrix}\label{dis}

The objective pursued when creating a typology of life course health trajectories is to categorise patients such that patients belonging to the same group share similar trajectories relatively to patients belonging to other groups. This process implies therefore the use of a dissimilarity/distance metric to obtain pairwise patients' profile dissimilarities to be stored in a distance matrix or \emph{dissimilarity index matrix}. Based on this matrix, a clustering method can be used to finally create a typology. 

As \emph{multiple state analysis} deals with state indicators, the relationship between two patients can be summarised using a 2 $\times$ 2 contingency table $[(p,r),(s,q)]$,% $s^i$ and $s^j$ being the time indicator of a given long term condition, $s^{i,j}$

\begin{enumerate}[(i)]
 \item $q$ being the number of matching time units where both patients are affected by a given long term condition,
 \item $p$ being the number of matching time units where both patients are free from the considered long term condition,
 \item $s$ and $r$ being the number of matching time units where both patients are in different states of health regarding the specified long term condition and,
 \item $t$ being the length of the sequence.
\end{enumerate}

In this setting, the dissimilarity between profiles $x^i$ and $x^j$ can be expressed as:

\begin{equation*}
\begin{cases}
 d(x^i,x^j)  =  \frac{ r + s }{ q + r + s + p }, \\
  q + r + s + p  =  t
\end{cases}
\end{equation*}

Considering states of illness as more informative than healthy states, omitting $p$, the number of negative matches in the denominator, leads to the Jaccard dissimilarity index \citep{Jaccard1901}:

$$ d(x^i,x^j) = \frac{ r + s }{ q + r + s } = 1 - \frac{ q }{ q + r + s }. $$

Since $ r + s = t - p - q$, we can rewrite the above relation as follows:  

$$ d(x^i,x^j) =  1 - \frac{ q }{ t - p }. $$

A composite analogue to the Jaccard dissimilarity index can be derived as:

\begin{equation}\label{CJDI}
\begin{cases*}
d(M^i,M^j) = 1 - \frac{ Q }{ t^* - P },\\
Q = \sum_{l=1}^k q_l,\\
P = \sum_{l=1}^k p_l,\\
t^* = kt
\end{cases*}
\end{equation}
 
i.e. $Q$ and $P$ are the sum of $q$ and $p$ over all considered long term conditions. 

~\\

Using the matrix notations (\ref{mat}) from \ref{sm} and setting $ \bar{\textrm{C}^i} $ as the $t \times t$ diagonal matrix with
$\bar{\textrm{c}^i}$ as diagonal entries, the censored quantities defined above can be computed as:

\begin{eqnarray*}
 Q &=& \textrm{tr} \left( \left( \textrm{M}^i{'}  \bar{\textrm{C}^i} \bar{\textrm{C}^j} \right)
 \left(\textrm{M}^j{'} \bar{\textrm{C}^i} \bar{\textrm{C}^j} \right)' \right)\\
 P &=&   \textrm{tr} \left(
 \left( \left( \textrm{M}^i - \mathds{1} \right)' \bar{\textrm{C}^i} \bar{\textrm{C}^j} \right)
 \left( \left( \textrm{M}^j - \mathds{1} \right)'  \bar{\textrm{C}^i} \bar{\textrm{C}^j} \right)' \right) \\ 
 t^* &=&  \textrm{tr}\left(
        \left( \mathds{1}' \bar{\textrm{C}^i} \bar{\textrm{C}^j}  \right)
        \left( \mathds{1}' \bar{\textrm{C}^i} \bar{\textrm{C}^j}  \right)'
        \right),
       \end{eqnarray*}

where $\textrm{tr}$ denotes the trace operator and $\mathds{1}$ is the $t \times k$ matrix with all entries set to 1.

Using this notation, we can translate the cases of two patients as described above. Patient $i$ has a record of hypertension ($hyp$) at 51 and a record of diabetes ($dm$) at 52, whereas patient $j$ has a record of diabetes at 51 and a record of hypertension at 52. 

Simple sequences of patients $i$ and $j$ would be:

\begin{equation*}
{x^i} =
\begin{blockarray}{cc}
 &   \\
\begin{block}{c( c )}
50 & 0  \\
51 & hyp  \\
52 & dm  \\
53 & dm \\
\vdots & \vdots  \\
60 & dm \\
\end{block}
\end{blockarray}\;,\:\: 
{x^j} =
\begin{blockarray}{cc}
 &   \\
\begin{block}{c( c )}
50 & 0  \\
51 & dm  \\
52 & hyp  \\
53 & hyp \\
\vdots & \vdots  \\
60 & hyp \\
\end{block}
\end{blockarray}\;,
\end{equation*}

whereas the state matrices of patients $i$ and $j$ would be: 

\begin{equation*}
  \mathbf{m^i}=
\begin{blockarray}{cccccc}
& dm & hyp \\
\begin{block}{c(ccccc)}
  50 & 0 & 0   \\
  51 & 0 & 1   \\
  52 & 1 & 1   \\
  53 & 1 & 1   \\
& \vdots & \vdots \\
  60 & 1 & 1   \\
\end{block}
\end{blockarray}\;,\:\:
  \mathbf{m^j}=
\begin{blockarray}{cccccc}
& dm & hyp \\
\begin{block}{c(ccccc)}
  50 & 0 & 0   \\
  51 & 1 & 0   \\
  52 & 1 & 1   \\
  53 & 1 & 1   \\
& \vdots & \vdots \\
  60 & 1 & 1   \\
\end{block}
\end{blockarray}\;.
\end{equation*}

The Jaccard dissimilarity index computed from the simple sequences of patients $i$ and $j$ equals 1 (its maximum). It equals $\frac{1}{10}$ using the state matrix notation and the composite Jaccard index proposed in (\ref{CJDI}), which is what we expect for these patients with similar health trajectories.

\subsubsection{Clustering method}

Although \emph{multiple state analysis} is not restricted to a specific clustering procedure, we have used in this paper the Ward's hierarchical clustering method \citep{ward1963hierarchical}. At the starting point of this procedure, each instance is considered as a cluster of its own, then clusters are recursively merged  such that the resulting cluster structure presents the minimum cost in terms of the \emph{within clusters} variance. This objective is reached by minimizing

\begin{equation*}
 W_{i,j} = \frac{n_i n_j}{n_i + n_j} \; d(c_i,c_j)^2
\end{equation*}

over $i$ and $j$, $N_i$ being the cardinality of cluster $i$ and $d(c_i,c_j)$ the Euclidean distance between centroids of clusters $i$ and $j$. Finally, dissimilarities are updated at each step following the Lance-Williams formula \citep{lance1967general}:

\begin{eqnarray*}
 d(c_i \cup c_j , c_k) & = & \alpha_i \; d(c_i,c_k) + \alpha_j \; d(c_j,c_k) + \beta \; d(c_i,c_j), \\
&~&
 \begin{cases}
\alpha_i = \frac{n_i + n_k}{n_i + n_j + n_k} \\
\alpha_j = \frac{n_j + n_k}{n_i + n_j + n_k} \\
\beta\:\:    =    \frac{n_k}{n_i + n_j + n_k}\;.
\end{cases}
\end{eqnarray*}

This formula allows the use of the Ward method in the general setting where $d(.)$ does not denote a \emph{proper} Euclidean distance. Following the Lance-Williams formula, the updated dissimilarity between merged clusters $i$ and $j$ ($c_i \cup c_j$) and any other clusters $k$ ($c_k$) is the simple linear combination of individual dissimilarities between involved clusters.
 
The result of this process is a dendrogram, representing the nested groups of patients and similarity levels at which clusters merged. Clustering of the data is obtained by cutting the dendrogram to get a workable number of clusters.

In the next section, we present an application of \emph{multiple state analysis} using electronic health records of patients with myocardial infarction and other comorbidity. We show subsequently how multivariate logistic regressions can be used to evaluate associations between clusters and socio-demographic variables and other risk factors.  

\section{Multiple State Analysis of multimorbidity associated to myocardial infarction}

\subsection{ Patients } 
\subsubsection{ Primary care registry }

In an application of the proposed method to real world data, we considered electronic health records 33 diseases and long term conditions in adult patients aged over 18 and registered in 41 general practices in south London between April 2005 and April 2021. The data were provided by the Lambeth DataNet. 
Recorded long term conditions were: cancers and neoplasms (1),
cardiovascular diseases (stroke (2), heart failure (3), peripheral vascular disease (4), atrial fibrillation (5), hypertension (6), myocardial infarction (7) and transient ischemic attack (8)),
infectious diseases	(viral hepatitis B \& C (9), Hiv/aids (10)),
inflammatory diseases / diseases of the immune system (rheumatoid arthritis (11) and inflamatory bowel disease (IBD) (12), lupus (13)), kidney diseases	(chronic kidney disease grade 3 to 5 (CKD 3-5) (14)),
liver diseases (15),
mental health conditions (anxiety disorders (16), depression (17), alcohol dependence (18), serious mental illness (19) and substance dependency (20),  learning disabilities (21)),
metabolic / endocrine diseases	(diabetes (22) and morbid obesity (23)), musculoskeletal conditions (osteoarthritis (24) and osteoporosis (25)),
Sickle-Cell Anaemia (26),
neurological conditions (parkinson's (27), multiple sclerosis (28), epilepsy (29) and dementia (30)), respiratory diseases (asthma (31) and COPD (32)) and chronic pain (33).
Patient electronic files includes the date at which any of the above long term conditions was ever recorded. Incident cases were defined as the first ever diagnostic record.

Of note some early diseases such as asthma were traced back and added to individual files outside patients follow-up interval. 

%\subsubsection{Myocardial Infarction}
%Myocardial Infarction (MI), leading to ``heart attack," is one of the leading causes of death in high-income countries \citep{roth2015global}. Myocardial infarction is caused by decreased or complete cessation of blood flow in the myocardium and results in irreversible damage to the heart muscle \citep{jaffe2013third}. Most of the time, myocardial infarction is due to underlying coronary artery disease \citep{Libby2013}. Well-known modifiable risk factors associated to coronary artery disease and myocardial infarction are smoking, abnormal blood lipid profile, hypertension, diabetes mellitus, abdominal obesity, psycho-social factors (depression, global stress, financial stress, and life events including marital separation, job loss, and family conflicts), diet,  physical activity and alcohol consumption (protective) \citep{Yusuf2004,Anand2008}. Some non-modifiable risk factors for myocardial infarction include advanced age, male gender (males tend to have myocardial infarction earlier in life) and genetics \citep{Anand2008, Nielsen2013}. 

\subsection{ Method }

The analysis was conducted on patients showing a record of myocardial infarction, according to the 3 steps described in \ref{MSA}: i) arrange patients individual records into multiple time-to-event endpoints stacked in individual patients' state matrices and censoring indicators,  ii) compute pairwise patients' dissimilarities on individual state matrices (and censoring indicators) and apply a clustering method, and iii) define a typology.

\begin{figure}[!ht]
\centering
\makebox{\includegraphics[page=1,scale = .4]{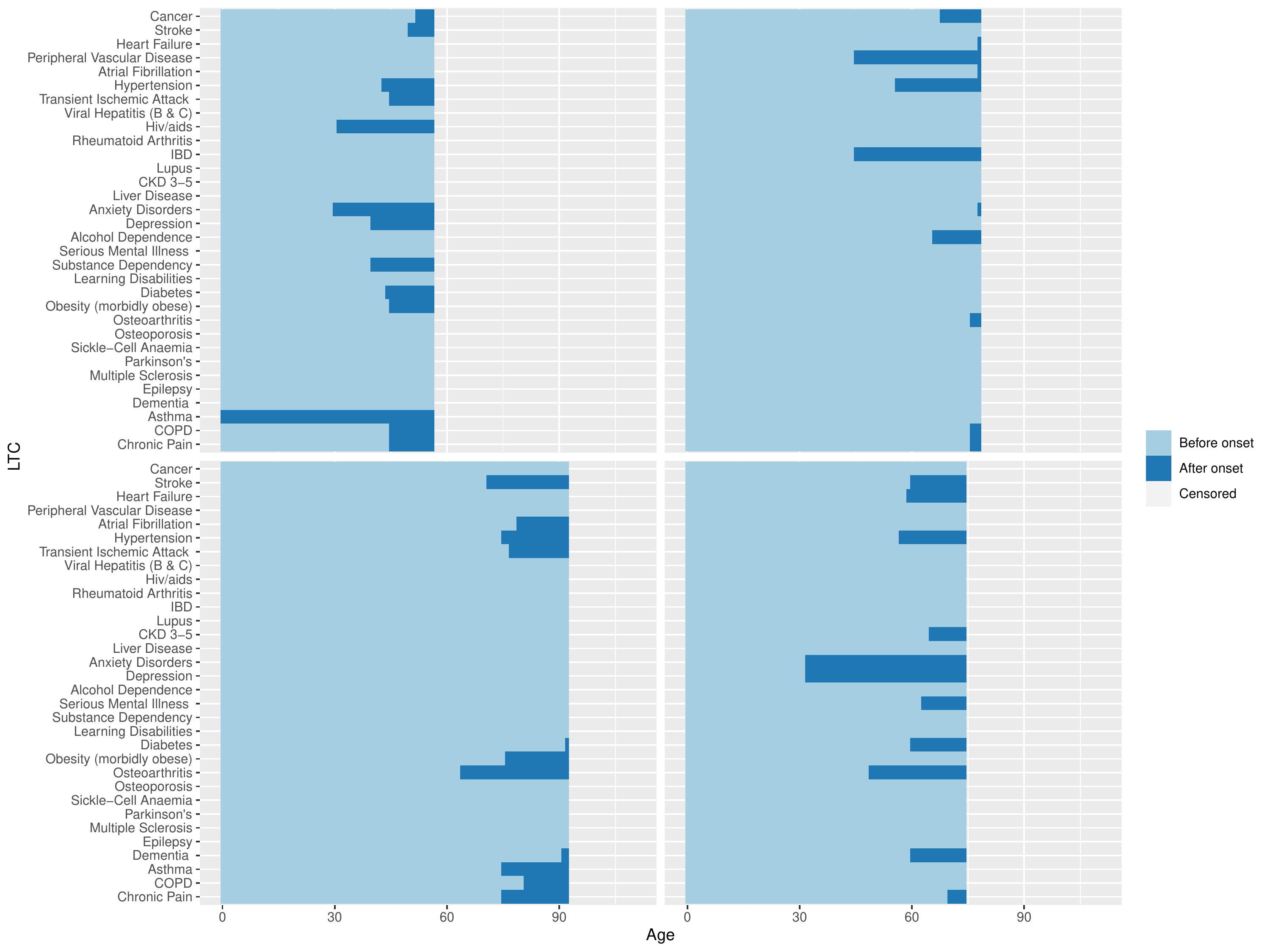}}
\caption{Example of life course patients' history (state matrices): status of 33 long term conditions are considered from 0 to 104 years old}
\label{fig:2}
\end{figure}

State matrices were computed considering the records of the 33 diseases and long term conditions enumerated above. Figure \ref{fig:2} represents example of state matrices from 4 patients randomly samples from the cohort.

Pairwise dissimilarities between patients were computed using the composite Jaccard dissimilarity index \citep{Jaccard1901} as described in \ref{dis}. The Jaccard coefficient, also referred to as the binary metric, is a measure of dissimilarity between binary samples, such as state indicator. It can be interpreted, for a given long term condition, as the amount of time two patients are simultaneously in a the same state (in term of patient's age) divided by the the amount of time these patients are in different states. Finally, agglomerative hierarchical clustering was computed using the Ward’s method and Hubert’s C index was used as an indicator in choosing a relevant size of typology in a convenient and workable range \citep{Hubert1976}.

\subsubsection{ Graph representation }
To better visualize and understand the link between onset of myocardial infarction and other associated long term conditions in defined clusters, we used a graph where conditions are represented by dots whose x-coordinate represents the median age at onset. On the y axis, dots are conveniently assigned to layers (levels), such that significant transitions are represented by edges oriented from a higher levels to a lower levels \citep{Sugiyama1981}. The number of layers in the graph can therefore be interpreted as the maximum number of significant transitions from the first event to the last event represented on the graph according to some threshold.

\subsubsection{ Statistical analysis of cluster trajectories }

In order to describe patients characteristics with respect to the proposed typology, socio-demographic variables, risk factors and diseases status indicators were crossed with cluster status indicators. All variables were binary or categorical except age at onset of myocardial infarction and the multimorbidity index. 

Socio-demographics, risk factors and long term condition indicators, were displayed as frequencies and percentage or median and interquartile range as appropriate. Associations between variables and clusters was tested using chi-squared test or Fisher exact test as appropriate, or using Kruskal–Wallis test for numeric variables (tables \ref{tab:t1} and \ref{tab:t2}). 

Multivariate association between clusters and socio-demographic factors / risk factors, and long term conditions was estimated using logistic regressions where cluster indicators were explained by tested variables. Results were displayed using heatmaps were non significant result, at the 5\% level, were omitted. 

\subsection{ Results }

From 856342 single patients for which electronic records were available, 5093 (0.6\%) had a record of myocardial infarction which corresponds to an incidence rate of 94.4 cases per 100,000 person-years. The median age at onset of myocardial infarction was 61. Among patients with a record of myocardial infarction, 68\% were male, 68\% were white and the majority of the population belonged to IMD quintile 1-2 (most deprived) (vs quintile 3-5 (less deprived)). The main comorbidities associated with myocardial infarction were hypertension (67\%) and  diabetes (40\%) and the median number of comorbidities associated with myocardial infarction was 5. The median follow-up was 9.6 years, 61\% of patients were censored or died at end of follow-up (39\%) (table \ref{tab:t1}). 

After pairwise patients' dissimilarity computation and the clustering, Hubert’s $C$ index was computed for a range of partition sizes. A 10 group partition, corresponding local minimum the $C$ index between 6 and 15 clusters was retained for this exploratory analysis.

\subsubsection{ Typology annotation }

\emph{Demographics and risk factors}

The 10 clusters were ordered by decreasing frequency and numbered from 1 to 10. Cluster 1 represents 1175 patients (23.1\%) and clusters 9 and 10 represent 218 (4.3\%) and 103 (2.0\%) patients respectively.

Table \ref{tab:t1} displays basic demographics and risk factors associated with clusters. Briefly, the median age at onset of myocardial infarction ranged from 47-48 in clusters 8 and 4 respectively, to 78 in cluster 2. The proportion of female patients ranged from 15.7\% in cluster 4 to 47.8\% in cluster 2. The proportion of White patients varied from 45.3\% in clusters 6 and 9 to 82.0\% in cluster 7, whereas the proportion of Black patients was 8.8\% in cluster 7 and reached 30.8\% in cluster 9, and the share of the Asian population ranged from 5.0\% in cluster 7 to 25.8\% in cluster 6. IMD-quintile index was not significantly different from cluster to cluster. Former or current smokers were over represented in cluster 7 (85\%), and  polymedication was more frequently observed in clusters 2, 3 and 6, and finally, the highest median number of comorbidities associated with myocardial infarction was observed in cluster 3 (median of 7).

\begin{figure}[!ht]
\centering
\makebox{\includegraphics[scale=.4, trim = 0cm 0cm 0cm 0cm, clip]{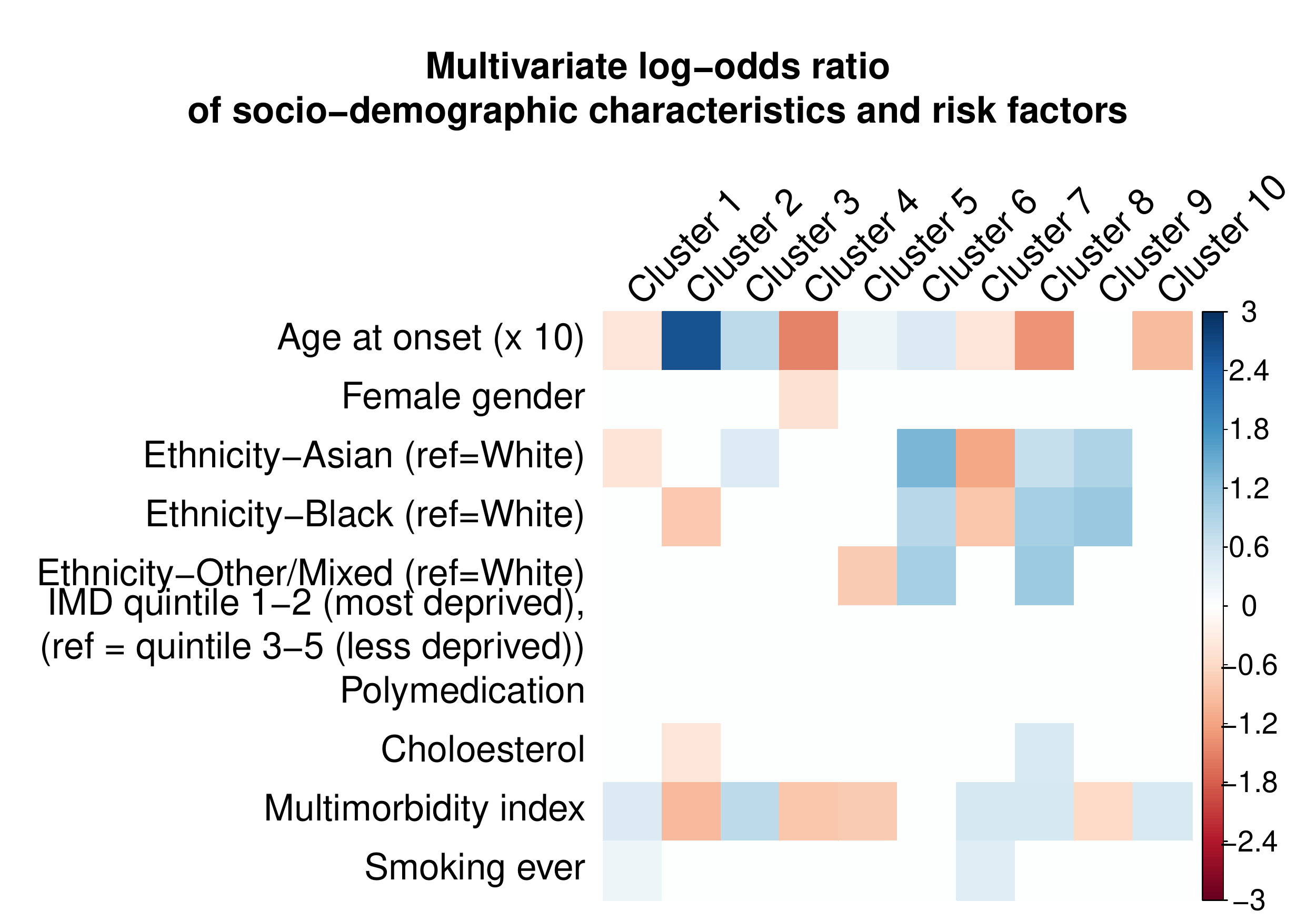}}
\caption{Log-odds ratio of socio-demographic variables and risk factors associated with clusters: values are derived from multivariate logistic regressions where cluster indicators are explained using displayed variables. Positive log-odds ratio show over representation of the corresponding population traits in a given cluster as compared to other patients showing a record of myocardial infarction. Only significant coefficients are displayed (at the 5\% level).}
\label{fig:3}
\end{figure}

Alternatively, figure \ref{fig:3} displays the multivariate log-odds ratio of demographic variables and risk factors associated with the different clusters.

\emph{Associated long term conditions}

The distribution of myocardial infarction comorbidities according to the different clusters is displayed in table \ref{tab:t2}: Among the most frequent comorbidities associated with myocardial infarction, including hypertension (67.1\%), diabetes (39.2\%), Osteoarthritis (29.8\%), depression (26.2\%) and asthma (15.5\%)), the proportion of hypertensive patients ranged from 40.8\% in cluster 10 to 100\% in cluster 9, diabetes proportion ranged from 23.8\% in clusters 2 to 99.5\% in cluster 6, osteoarthritis proportion ranged from 17.5\% in cluster 10 to 60.2\% in cluster 3, depression proportion ranged from 10.1 in cluster 5 and 6 / 10.2 in cluster 2 to 86.2\% in cluster 7, and asthma proportion ranged from 5.0\% in cluster 5 to 100\% in cluster 10.

\begin{figure}[!ht]
\centering
\includegraphics[scale= 1]{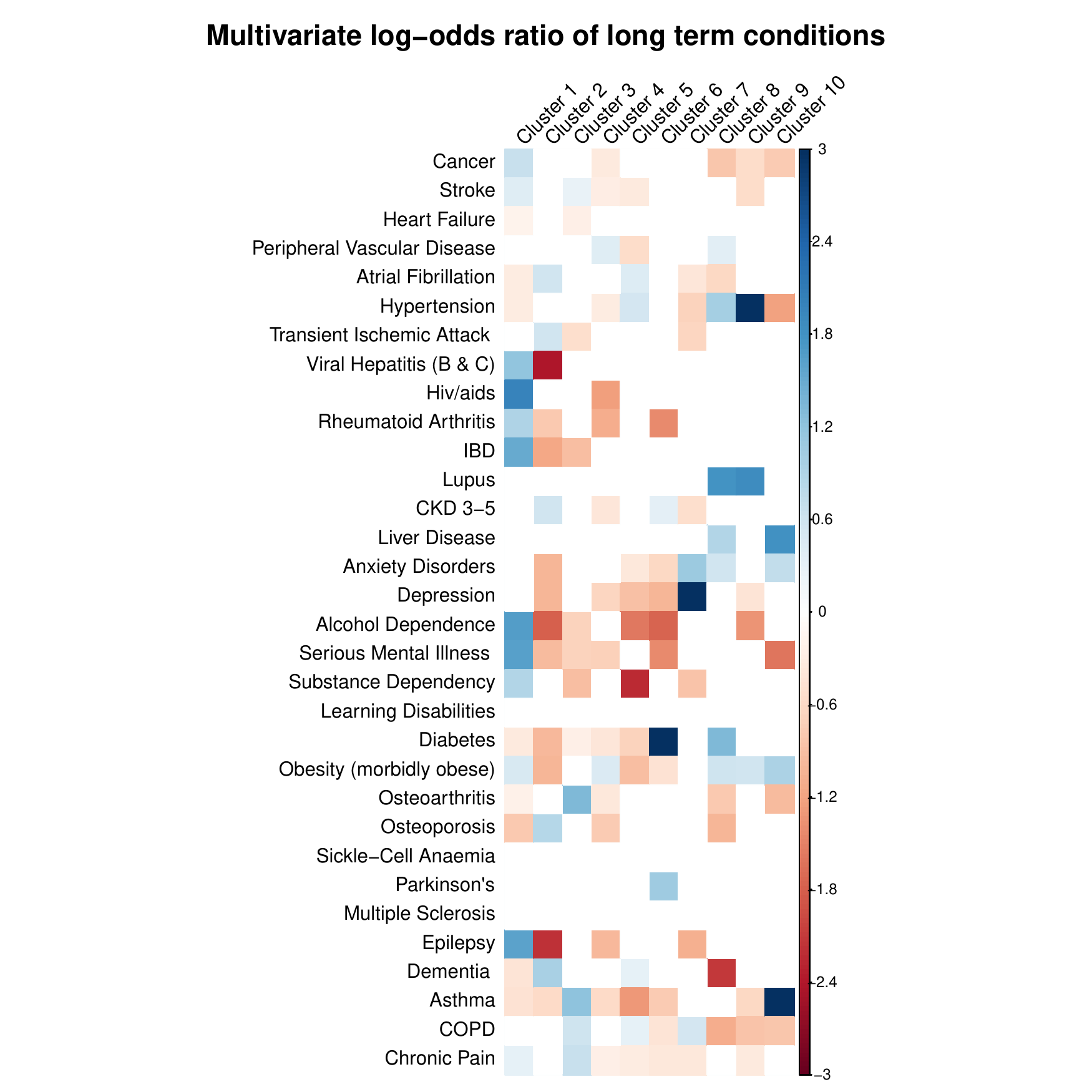}
\caption{Log-odds ratio of the 32 long term conditions associated with clusters: values are derived from multivariate logistic regressions where cluster indicators are explained by displayed variables. Positive log-odds ratio show over representation of the corresponding population traits in a given cluster as compared to other patients showing a record of myocardial infarction. Only significant coefficients are displayed (at the 5\% level).}
\label{fig:4}
\end{figure}

\begin{figure}[!ht]
\centering
\makebox{\includegraphics[page=1 , width=1\textwidth ,  trim = 0cm 0cm 0cm 0cm , clip]{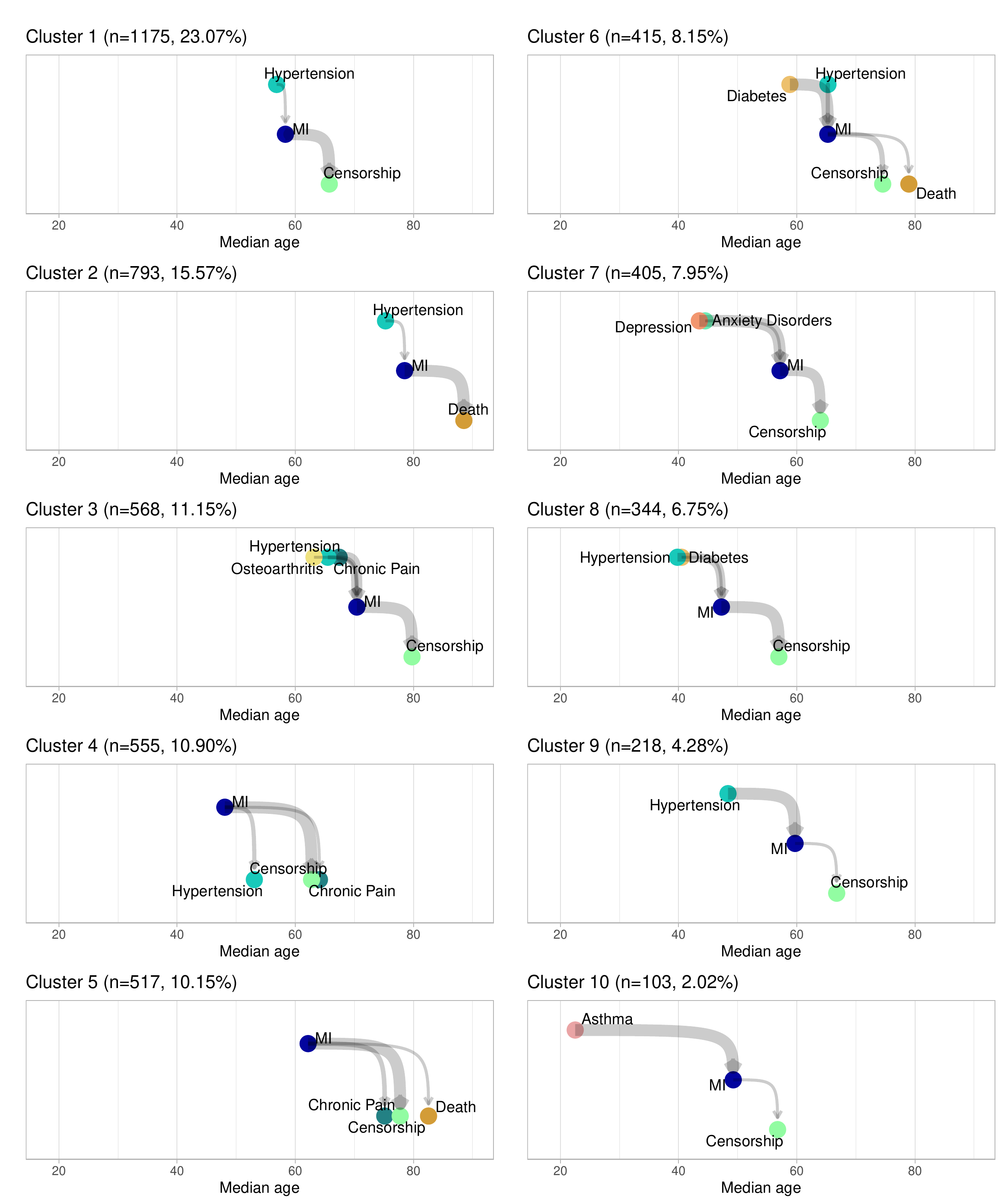}}
\caption{Graphical representation of clusters: long term conditions are represented by colored dots. Dots x-coordinate represents the median age at onset of the given long term condition. Dots are presented on the the y axis according to layers, such that no significant transitions exist between dots from the same layer. Significant transitions are represented by edges oriented from a higher level to a lower levels.}
\label{fig:5}
\end{figure}

Figure \ref{fig:4} displays the multivariate log-odds ratio of long term conditions associated with clusters, and finally a graphical representations of the onset of myocardial infarction and associated long term conditions in proposed in figure \ref{fig:5}.

\subsection{ Clusters examination }

From tables \ref{tab:t1} and \ref{tab:t2}, and results of the above multivariate statistical analysis on patients' socio-demographic variables, risk factors and associated long term conditions, it is now possible to better characterise the resulting typology of myocardial infarction. 

Hypertension is the most frequent long term condition associated with myocardial infarction in this study. It precedes onset of myocardial infarction in 2/3 of hypertensive patients. This pattern is the main characteristic of cluster 1 and 2 (accounting for 23.1\% and 15.6\% of analyzed patients respectively). Median age at onset of myocardial infarction significantly differs in each cluster, from 57.6 year in cluster 1 to 78.2 year in cluster 2 . Cluster 1 is also characterized by a high multimorbidity index with a median of 6. Comorbidities observed in cluster 1 include infectious diseases (HIV, B \& C viral hepatitis), inflammatory conditions (IBD, rheumatoid arthritis), mental illness, substance dependency (including alcohol and smoking) and stroke. Comorbidities observed in cluster 2 are linked to higher age, they include dementia and osteoporosis but also cardiovascular conditions such as atrial fibrillation and transient ischemic attack.

Cluster 3 (11.2\% of patients) is also characterized by a high multimorbidity index (median of 7), and a history of asthma, COPD and osteoarthritis.

Cluster 4 (10.9\% of patients) is singular: most of patients in this group are younger male patients with a median of 48.1 year at onset of myocardial infarction. Patients in this cluster have also the lowest multimorbidity index (median of 4).

The structure of cluster 5 (10.2\% of patients) is similar to cluster 4 with a low multimorbidity index (median of 5). However, patients in this cluster are significantly older at onset of myocardial infarction (median age of 62.2) and significantly more affected by hypertension (75.6\% vs 58.7\%) and COPD.

Cluster 6 and 8, accounting for 8.1\% and 6.8\% of analyzed patients respectively, are characterised by diverse metabolic disorders including diabetes (99.5\% and 71.2\% of patients respectively). Cluster 8 has a higher mulimorbidity index (median of 6). Along with diabetes, the main conditions associated with myocardial infarction in this cluster are cardio-vascular conditions (hypertension (83.7\%) and peripheral vascular disease), anxiety disorders and morbid obesity. Importantly, if both clusters are also characterised by a higher proportion of non-White patients, cluster 8 is the youngest group of patients with a median age at onset of myocardial infarction of 47.2 (vs 65.2 in cluster 6).

White patients are over represented in cluster 7 (82.0\%). This cluster, accounting for 8.0\% of analyzed patients, is also characterised by mental health conditions (including depression (86.2\%) and anxiety disorders), COPD, and the highest proportion of smokers or former smoker (85.4\%).

Like clusters 6 and 8, non-White patients are over represented in cluster 9 (accounting for 4.3\% of analyzed patients). This cluster is also characterised by a high rate of morbid obese patients and more strikingly by hypertension as 100\% of patient in this cluster suffer from this long term condition.

Finally, cluster 10 is characterized by a young population at onset of myocardial infarction (49.2 year) with a high multimorbidity index (median of 6). Along with asthma which affects all patients in this cluster, the other comorbidities associated with myocardial infarction in this cluster are: morbid obesity, liver disease and anxiety disorder. 

\section{ Discussion and concluding remarks }
 
We have presented in this paper a novel method aimed to analysing multiple time-to-event data denoted as \emph{multiple state analysis} with application to the analysis of life-course health trajectories. This method relates to the sequence analysis developed in the field of the social sciences, and more particularly to Pollock's multivariate extension of sequence analysis, where multiple aspects of the social experience, such as marital and employment status, are jointly analysed using a \emph{multiple sequences analysis} strategy. Before pairwise subject dissimilarities computation and its subsequent clustering phase, the proposed \emph{multiple state analysis} requires only a binary coding of individual data into states matrices and associated follow up indicator. From this, a typology of the main life course trajectories is derived.

In an application of \emph{multiple state analysis} to multimorbidity associated with myocardial infarction, we analysed 5093 unique individual patients experiencing up to 32 different commodities, aged from 0 to 104, with various follow-up times. Following our strategy, patients could be mapped back to a limited number of clusters associated with major life-long health trajectories that is, dominant sequences of long term conditions associated with myocardial infarction showing distinct demographic characteristics and levels of the main known myocardial infarction risk factors:
\begin{enumerate}[(a)]
 \item hypertension, which is globally the main driver of myocardial infarction, is not evenly distributed across clusters. Examination of cluster 9 reveals that myocardial infarction occurs in this cluster in the context of a longer and systematic background of hypertension, which is diagnosed with a median of 10 years before the onset of myocardial infarction in all patients in this cluster,
\item a similar observation can be made in clusters  6, 7 and 10 with diabetes, depression and asthma respectively.
Of note, all patients in cluster 10 have a history of asthma, which was diagnosed with a median of more than 26 years before the onset of myocardial infarction. This pattern is also seen in cluster 3, where 80.2\% of patients with myocardial infarction have a history of osteoarthritis and/or asthma,
\item in clusters 4 and 5, which are characterized by low multimorbidity indexes, myocardial infarction arises in a non-specific background. However, median age at onset of myocardial infarction is lower in cluster 4 as compared to clusters 5 (48.1 vs 62.2 respectively),
\item finally, our results highlight the link between ethnicity and driving long term conditions such as diabetes which is associated with Asian, Black and Mixed ethnicity in cluster 6 and 8, and hypertension which is associated with Asian and Black ethnicity in cluster 9.  
\end{enumerate}

The ability to produce a limited number of distinct and meaningful profiles from arbitrary large populations, as presented in the proposed application of \emph{multiple state analysis}, has been denoted by Pollock as a powerful trait of his technique (\emph{multiple sequence analysis}) which may therefore be used as an end in itself \citep{Pollock2007}. 

Without loss of generality, the approach proposed in this paper can be seen as an extension of Pollock's multiple sequence analysis when patients (or any subject of interest) may experience an arbitrary number of events for which time of occurrence is of interest. Of note, the method is not constrained by the record of rare events and it allows for right or interval censoring. Finally, its implementation is straightforward using the basic libraries of popular statistical software. 

\emph{Multiple state analysis} is particularly appropriate to the analysis of multiple comorbidities as this phenomenon is characterized by the accumulation over time of potentially many long term conditions such as diabetes or hypertension as well as acute events such as myocardial infarction or stroke. In contrast to situations where transition between exclusive states is of interest, such as changes in the marital status in the analysis of social trajectories, in the epidemiological setting of multimorbidity, patients' health trajectories are characterised by different levels of multimorbidity over life courses. In this setting, the simple record of events, and its resulting sequences, analysed as sequences of exclusive events, may lead to irrelevant conclusions.   

Examination of the typology created using \emph{multiple state analysis} shows that resulting groups of patients are discriminated in terms of the main myocardial infarction driving long term conditions but also with respect to major demographics variables including age, sex and ethnicity, and common risk factors such as cholesterol or smoking status. Assuming that early life exposures as well as some latter behaviors / exposures are implicitly apprehended through the continuous record of acute diseases and long term conditions, theses results support the environmental approach hypothesis of life course epidemiology whether it focuses on early life exposure or later behaviors / exposures. Our results also implicitly support a genotype / environment hypothesis to myocardial infarction, depending on the extent to which long term conditions driving the occurrence of myocardial infarction are influenced by gene / environment exposure.

From a public health perspective however, the genotype / environment balance origin of medical records history is not of prime importance as public health policies would rather focus on modifiable risk factors of diseases. To this end, \emph{multiple state analysis} is a particularly relevant tool as it allows stratification of at-risk populations in terms of the main demographics variable as well as the main driving long term conditions allowing tailored communication according to target population. 

To conclude, the proposed approach to multiple time-to-event data denoted here as \emph{multiple state analysis} is a versatile tool, comparable in its objective and results to the simple or multiple analysis of sequence data, but adapted to the situation where the times to potentially many non exclusive events need to be analysed in a multidimensional manner as opposed to the simple analysis of the resulting sequences of events. Alike the simple or multiple sequence analysis, the aim of \emph{multiple state analysis} is the development of exploratory typologies allowing to uncover broad distinctions between majors life course trajectories, without focusing on a single event or point in time. As such, results of \emph{multiple state analysis} may be valuable for further modeling, to guide public health decision making or as an end in itself \citep{Pollock2007}.     

\section*{ Acknowledgements }
We would like to thank Mark Ashworth, Stevo Durbaba, Xiaohui Sun, and Mohammad Yadegarfar for their help in data extraction and approvals. Approval for the analysis of fully anonymized data was granted by Lambeth DataNet Clinical Commissioning Group and Information Governance Steering Group. 
This project is funded by King’s Health Partners / Guy’s and St Thomas Charity ‘MLTC Challenge Fund’ (grant number EIC180702) and support from the National Institute for Health and Care Research (NIHR) under its Programme Grants for Applied Research (NIHR202339) and the NIHR Applied Research Collaboration (ARC) South London at King’s College Hospital NHS Foundation Trust. The views expressed are those of the authors and not necessarily those of the NIHR or the Department of Health and Social Care.

     % use whatever margins you want for left, right, top and bottom.

\newgeometry{top=0cm, bottom=0cm, left=2cm , right=0cm}
\begin{landscape}\begingroup\fontsize{8}{12}\selectfont
\thispagestyle{empty}
\begin{longtable}[t]{llllllllllllll}
\caption{\label{tab:t1} \normalsize{Distribution of socio-demographic variables and risk factors associated with myocardial infarction according the clusters}}\\
\toprule
Cluster \# &  & 1 & 2 & 3 & 4 & 5 & 6 & 7 & 8 & 9 & 10 & Total & p\\
\midrule
\endfirsthead
\caption[]{Distribution of socio-demographic variables and risk factors associated with myocardial infarction according the clusters \textit{(continued)}}\\
\toprule
Cluster \# &  & 1 & 2 & 3 & 4 & 5 & 6 & 7 & 8 & 9 & 10 & Total & p\\
\midrule
\endhead
\midrule
\multicolumn{14}{r@{}}{\textit{(Continued on Next Page...)}}\
\endfoot
\bottomrule
\multicolumn{14}{l}{\rule{0pt}{1em}\textsuperscript{$\dagger$} IMD quitile 1 (most deprived)-2 vs 3-5 (less deprived)}\\
\endlastfoot
\cellcolor{gray!6}{Sex} & \cellcolor{gray!6}{Female} & \cellcolor{gray!6}{320 (27.2)} & \cellcolor{gray!6}{379 (47.8)} & \cellcolor{gray!6}{257 (45.2)} & \cellcolor{gray!6}{87 (15.7)} & \cellcolor{gray!6}{136 (26.3)} & \cellcolor{gray!6}{135 (32.5)} & \cellcolor{gray!6}{135 (33.3)} & \cellcolor{gray!6}{94 (27.3)} & \cellcolor{gray!6}{73 (33.5)} & \cellcolor{gray!6}{24 (23.3)} & \cellcolor{gray!6}{1640 (32.2)} & \cellcolor{gray!6}{<0.001}\\
Age at onset & Median (IQR) & 57.6 (14.6) & 78.2 (12.4) & 70.2 (12.2) & 48.1 (7.9) & 62.2 (8.3) & 65.2 (12.8) & 57.2 (15.4) & 47.2 (13.0) & 59.7 (12.6) & 49.2 (14.6) & 61.2 (19.9) & <0.001\\
\cellcolor{gray!6}{Ethnicity} & \cellcolor{gray!6}{White} & \cellcolor{gray!6}{617 (71.2)} & \cellcolor{gray!6}{416 (76.8)} & \cellcolor{gray!6}{282 (68.4)} & \cellcolor{gray!6}{286 (68.4)} & \cellcolor{gray!6}{270 (73.2)} & \cellcolor{gray!6}{135 (45.3)} & \cellcolor{gray!6}{260 (82.0)} & \cellcolor{gray!6}{123 (47.7)} & \cellcolor{gray!6}{72 (45.3)} & \cellcolor{gray!6}{49 (62.0)} & \cellcolor{gray!6}{2510 (67.5)} & \cellcolor{gray!6}{<0.001}\\
 & Asian & 77 (8.9) & 46 (8.5) & 55 (13.3) & 51 (12.2) & 38 (10.3) & 77 (25.8) & 16 (5.0) & 39 (15.1) & 31 (19.5) & 9 (11.4) & 439 (11.8) & \\
\cellcolor{gray!6}{} & \cellcolor{gray!6}{Black} & \cellcolor{gray!6}{133 (15.3)} & \cellcolor{gray!6}{59 (10.9)} & \cellcolor{gray!6}{66 (16.0)} & \cellcolor{gray!6}{58 (13.9)} & \cellcolor{gray!6}{52 (14.1)} & \cellcolor{gray!6}{65 (21.8)} & \cellcolor{gray!6}{28 (8.8)} & \cellcolor{gray!6}{69 (26.7)} & \cellcolor{gray!6}{49 (30.8)} & \cellcolor{gray!6}{14 (17.7)} & \cellcolor{gray!6}{593 (15.9)} & \cellcolor{gray!6}{}\\
 & Other & 40 (4.6) & 21 (3.9) & 9 (2.2) & 23 (5.5) & 9 (2.4) & 21 (7.0) & 13 (4.1) & 27 (10.5) & 7 (4.4) & 7 (8.9) & 177 (4.8) & \\
\cellcolor{gray!6}{IMD quintile \textsuperscript{$\dagger$}} & \cellcolor{gray!6}{1-2} & \cellcolor{gray!6}{786 (66.9)} & \cellcolor{gray!6}{527 (66.5)} & \cellcolor{gray!6}{389 (68.5)} & \cellcolor{gray!6}{344 (62.0)} & \cellcolor{gray!6}{321 (62.1)} & \cellcolor{gray!6}{269 (64.8)} & \cellcolor{gray!6}{281 (69.4)} & \cellcolor{gray!6}{243 (70.6)} & \cellcolor{gray!6}{138 (63.3)} & \cellcolor{gray!6}{70 (68.0)} & \cellcolor{gray!6}{3368 (66.1)} & \cellcolor{gray!6}{0.065}\\
Cholesterol & yes & 860 (75.0) & 528 (69.9) & 445 (79.0) & 385 (72.1) & 379 (74.3) & 292 (72.5) & 318 (78.9) & 271 (79.5) & 157 (72.0) & 68 (67.3) & 3703 (74.4) & 0.001\\
\cellcolor{gray!6}{Smoking ever} & \cellcolor{gray!6}{yes} & \cellcolor{gray!6}{941 (82.0)} & \cellcolor{gray!6}{474 (62.8)} & \cellcolor{gray!6}{405 (71.9)} & \cellcolor{gray!6}{430 (80.5)} & \cellcolor{gray!6}{377 (73.9)} & \cellcolor{gray!6}{280 (69.5)} & \cellcolor{gray!6}{344 (85.4)} & \cellcolor{gray!6}{271 (79.5)} & \cellcolor{gray!6}{144 (66.1)} & \cellcolor{gray!6}{79 (78.2)} & \cellcolor{gray!6}{3745 (75.3)} & \cellcolor{gray!6}{<0.001}\\
Polymedication & yes & 426 (36.8) & 337 (44.2) & 244 (43.3) & 149 (27.6) & 170 (33.9) & 177 (44.0) & 143 (35.9) & 122 (35.7) & 69 (31.9) & 30 (29.1) & 1867 (37.4) & <0.001\\
\cellcolor{gray!6}{Multimorbidity index} & \cellcolor{gray!6}{Median (IQR)} & \cellcolor{gray!6}{6.0 (3.0)} & \cellcolor{gray!6}{5.0 (3.0)} & \cellcolor{gray!6}{7.0 (3.0)} & \cellcolor{gray!6}{4.0 (3.0)} & \cellcolor{gray!6}{5.0 (4.0)} & \cellcolor{gray!6}{5.0 (3.0)} & \cellcolor{gray!6}{6.0 (4.0)} & \cellcolor{gray!6}{6.0 (4.0)} & \cellcolor{gray!6}{5.0 (4.0)} & \cellcolor{gray!6}{5.0 (3.0)} & \cellcolor{gray!6}{5.0 (3.0)} & \cellcolor{gray!6}{<0.001}\\
Status & Dead & 389 (33.1) & 518 (65.3) & 244 (43.0) & 170 (30.6) & 222 (42.9) & 187 (45.1) & 101 (24.9) & 90 (26.2) & 59 (27.1) & 21 (20.4) & 2001 (39.3) & <0.001\\
\cellcolor{gray!6}{Total N (\%)} & \cellcolor{gray!6}{} & \cellcolor{gray!6}{1175 (23.1)} & \cellcolor{gray!6}{793 (15.6)} & \cellcolor{gray!6}{568 (11.2)} & \cellcolor{gray!6}{555 (10.9)} & \cellcolor{gray!6}{517 (10.2)} & \cellcolor{gray!6}{415 (8.1)} & \cellcolor{gray!6}{405 (8.0)} & \cellcolor{gray!6}{344 (6.8)} & \cellcolor{gray!6}{218 (4.3)} & \cellcolor{gray!6}{103 (2.0)} & \cellcolor{gray!6}{5093} & \cellcolor{gray!6}{}\\*
\end{longtable}
\endgroup{}

\end{landscape}

\break

\begin{landscape}\begingroup\fontsize{8}{12}\selectfont
\thispagestyle{empty}
\begin{longtable}[t]{lllllllllllll}
\caption{\label{tab:t2}Distribution of the 32 comorbidities associated with myocardial infarction according the clusters}\\
\toprule
Cluster \# & 1 & 2 & 3 & 4 & 5 & 6 & 7 & 8 & 9 & 10 & Total & p\\
\midrule
\endfirsthead
\caption[]{Distribution of the 32 comorbidities associated with myocardial infarction according the clusters \textit{(continued)}}\\
\toprule
Cluster \# & 1 & 2 & 3 & 4 & 5 & 6 & 7 & 8 & 9 & 10 & Total & p\\
\midrule
\endhead
\midrule
\multicolumn{13}{r@{}}{\textit{(Continued on Next Page...)}}\
\endfoot
\bottomrule
\multicolumn{13}{l}{\rule{0pt}{1em}\textsuperscript{$\dagger$} Inflammatory Bowel Disease}\\
\multicolumn{13}{l}{\rule{0pt}{1em}\textsuperscript{$\ddagger$} Chronic Obstructive Pulmonary Disease}\\
\multicolumn{13}{l}{\rule{0pt}{1em}\textsuperscript{$\S$} Chronic Kidney Disease stage 3 to stage 5}\\
\endlastfoot
\cellcolor{gray!6}{Cancer} & \cellcolor{gray!6}{273 (23.2)} & \cellcolor{gray!6}{141 (17.8)} & \cellcolor{gray!6}{107 (18.8)} & \cellcolor{gray!6}{71 (12.8)} & \cellcolor{gray!6}{96 (18.6)} & \cellcolor{gray!6}{69 (16.6)} & \cellcolor{gray!6}{61 (15.1)} & \cellcolor{gray!6}{28 (8.1)} & \cellcolor{gray!6}{23 (10.6)} & \cellcolor{gray!6}{11 (10.7)} & \cellcolor{gray!6}{880 (17.3)} & \cellcolor{gray!6}{<0.001}\\
Stroke & 235 (20.0) & 149 (18.8) & 112 (19.7) & 65 (11.7) & 71 (13.7) & 72 (17.3) & 57 (14.1) & 55 (16.0) & 26 (11.9) & 14 (13.6) & 856 (16.8) & <0.001\\
\cellcolor{gray!6}{Heart Failure} & \cellcolor{gray!6}{292 (24.9)} & \cellcolor{gray!6}{271 (34.2)} & \cellcolor{gray!6}{162 (28.5)} & \cellcolor{gray!6}{157 (28.3)} & \cellcolor{gray!6}{170 (32.9)} & \cellcolor{gray!6}{128 (30.8)} & \cellcolor{gray!6}{79 (19.5)} & \cellcolor{gray!6}{109 (31.7)} & \cellcolor{gray!6}{65 (29.8)} & \cellcolor{gray!6}{23 (22.3)} & \cellcolor{gray!6}{1456 (28.6)} & \cellcolor{gray!6}{<0.001}\\
Peripheral Vascular Disease & 141 (12.0) & 92 (11.6) & 68 (12.0) & 74 (13.3) & 38 (7.4) & 70 (16.9) & 45 (11.1) & 58 (16.9) & 26 (11.9) & 5 (4.9) & 617 (12.1) & <0.001\\
\cellcolor{gray!6}{Atrial Fibrillation} & \cellcolor{gray!6}{166 (14.1)} & \cellcolor{gray!6}{231 (29.1)} & \cellcolor{gray!6}{133 (23.4)} & \cellcolor{gray!6}{86 (15.5)} & \cellcolor{gray!6}{133 (25.7)} & \cellcolor{gray!6}{65 (15.7)} & \cellcolor{gray!6}{42 (10.4)} & \cellcolor{gray!6}{38 (11.0)} & \cellcolor{gray!6}{33 (15.1)} & \cellcolor{gray!6}{11 (10.7)} & \cellcolor{gray!6}{938 (18.4)} & \cellcolor{gray!6}{<0.001}\\
Hypertension  & 706 (60.1) & 562 (70.9) & 389 (68.5) & 326 (58.7) & 391 (75.6) & 290 (69.9) & 205 (50.6) & 288 (83.7) & 218 (100.0) & 42 (40.8) & 3417 (67.1) & <0.001\\
\cellcolor{gray!6}{Transient Ischemic Attack} & \cellcolor{gray!6}{78 (6.6)} & \cellcolor{gray!6}{89 (11.2)} & \cellcolor{gray!6}{32 (5.6)} & \cellcolor{gray!6}{30 (5.4)} & \cellcolor{gray!6}{35 (6.8)} & \cellcolor{gray!6}{26 (6.3)} & \cellcolor{gray!6}{17 (4.2)} & \cellcolor{gray!6}{21 (6.1)} & \cellcolor{gray!6}{19 (8.7)} & \cellcolor{gray!6}{4 (3.9)} & \cellcolor{gray!6}{351 (6.9)} & \cellcolor{gray!6}{<0.001}\\
Viral Hepatitis (B \& C) & 41 (3.5) & 1 (0.1) & 6 (1.1) & 8 (1.4) & 1 (0.2) & 4 (1.0) & 4 (1.0) & 5 (1.5) & 1 (0.5) & 1 (1.0) & 72 (1.4) & <0.001\\
\cellcolor{gray!6}{Hiv/aids} & \cellcolor{gray!6}{52 (4.4)} & \cellcolor{gray!6}{0 (0.0)} & \cellcolor{gray!6}{0 (0.0)} & \cellcolor{gray!6}{3 (0.5)} & \cellcolor{gray!6}{2 (0.4)} & \cellcolor{gray!6}{0 (0.0)} & \cellcolor{gray!6}{8 (2.0)} & \cellcolor{gray!6}{4 (1.2)} & \cellcolor{gray!6}{1 (0.5)} & \cellcolor{gray!6}{5 (4.9)} & \cellcolor{gray!6}{75 (1.5)} & \cellcolor{gray!6}{<0.001}\\
Rheumatoid Arthritis & 46 (3.9) & 12 (1.5) & 28 (4.9) & 4 (0.7) & 12 (2.3) & 3 (0.7) & 11 (2.7) & 7 (2.0) & 4 (1.8) & 5 (4.9) & 132 (2.6) & <0.001\\
\cellcolor{gray!6}{IBD \textsuperscript{$\dagger$}} & \cellcolor{gray!6}{45 (3.8)} & \cellcolor{gray!6}{6 (0.8)} & \cellcolor{gray!6}{8 (1.4)} & \cellcolor{gray!6}{9 (1.6)} & \cellcolor{gray!6}{9 (1.7)} & \cellcolor{gray!6}{1 (0.2)} & \cellcolor{gray!6}{4 (1.0)} & \cellcolor{gray!6}{3 (0.9)} & \cellcolor{gray!6}{1 (0.5)} & \cellcolor{gray!6}{4 (3.9)} & \cellcolor{gray!6}{90 (1.8)} & \cellcolor{gray!6}{<0.001}\\
Lupus & 6 (0.5) & 0 (0.0) & 2 (0.4) & 0 (0.0) & 0 (0.0) & 0 (0.0) & 2 (0.5) & 4 (1.2) & 2 (0.9) & 0 (0.0) & 16 (0.3) & 0.016\\
\cellcolor{gray!6}{CKD 3-5 \textsuperscript{$\ddagger$} } & \cellcolor{gray!6}{270 (23.0)} & \cellcolor{gray!6}{318 (40.1)} & \cellcolor{gray!6}{169 (29.8)} & \cellcolor{gray!6}{105 (18.9)} & \cellcolor{gray!6}{167 (32.3)} & \cellcolor{gray!6}{155 (37.3)} & \cellcolor{gray!6}{67 (16.5)} & \cellcolor{gray!6}{94 (27.3)} & \cellcolor{gray!6}{64 (29.4)} & \cellcolor{gray!6}{13 (12.6)} & \cellcolor{gray!6}{1422 (27.9)} & \cellcolor{gray!6}{<0.001}\\
Liver Disease & 40 (3.4) & 6 (0.8) & 5 (0.9) & 5 (0.9) & 5 (1.0) & 5 (1.2) & 4 (1.0) & 12 (3.5) & 3 (1.4) & 3 (2.9) & 88 (1.7) & <0.001\\
\cellcolor{gray!6}{Anxiety Disorders } & \cellcolor{gray!6}{299 (25.4)} & \cellcolor{gray!6}{64 (8.1)} & \cellcolor{gray!6}{149 (26.2)} & \cellcolor{gray!6}{81 (14.6)} & \cellcolor{gray!6}{58 (11.2)} & \cellcolor{gray!6}{33 (8.0)} & \cellcolor{gray!6}{239 (59.0)} & \cellcolor{gray!6}{100 (29.1)} & \cellcolor{gray!6}{24 (11.0)} & \cellcolor{gray!6}{37 (35.9)} & \cellcolor{gray!6}{1084 (21.3)} & \cellcolor{gray!6}{<0.001}\\
Depression & 393 (33.4) & 81 (10.2) & 160 (28.2) & 78 (14.1) & 52 (10.1) & 42 (10.1) & 349 (86.2) & 114 (33.1) & 31 (14.2) & 35 (34.0) & 1335 (26.2) & <0.001\\
\cellcolor{gray!6}{Alcohol Dependence} & \cellcolor{gray!6}{246 (20.9)} & \cellcolor{gray!6}{10 (1.3)} & \cellcolor{gray!6}{25 (4.4)} & \cellcolor{gray!6}{37 (6.7)} & \cellcolor{gray!6}{8 (1.5)} & \cellcolor{gray!6}{5 (1.2)} & \cellcolor{gray!6}{49 (12.1)} & \cellcolor{gray!6}{23 (6.7)} & \cellcolor{gray!6}{4 (1.8)} & \cellcolor{gray!6}{8 (7.8)} & \cellcolor{gray!6}{415 (8.1)} & \cellcolor{gray!6}{<0.001}\\
Serious Mental Illness & 113 (9.6) & 8 (1.0) & 10 (1.8) & 9 (1.6) & 5 (1.0) & 4 (1.0) & 25 (6.2) & 13 (3.8) & 2 (0.9) & 2 (1.9) & 191 (3.8) & <0.001\\
\cellcolor{gray!6}{Substance Dependency} & \cellcolor{gray!6}{105 (8.9)} & \cellcolor{gray!6}{11 (1.4)} & \cellcolor{gray!6}{9 (1.6)} & \cellcolor{gray!6}{12 (2.2)} & \cellcolor{gray!6}{1 (0.2)} & \cellcolor{gray!6}{3 (0.7)} & \cellcolor{gray!6}{18 (4.4)} & \cellcolor{gray!6}{13 (3.8)} & \cellcolor{gray!6}{2 (0.9)} & \cellcolor{gray!6}{7 (6.8)} & \cellcolor{gray!6}{181 (3.6)} & \cellcolor{gray!6}{<0.001}\\
Learning Disabilities  & 14 (1.2) & 0 (0.0) & 0 (0.0) & 4 (0.7) & 1 (0.2) & 3 (0.7) & 4 (1.0) & 0 (0.0) & 0 (0.0) & 0 (0.0) & 26 (0.5) & 0.002\\
\cellcolor{gray!6}{Diabetes} & \cellcolor{gray!6}{378 (32.2)} & \cellcolor{gray!6}{189 (23.8)} & \cellcolor{gray!6}{201 (35.4)} & \cellcolor{gray!6}{178 (32.1)} & \cellcolor{gray!6}{133 (25.7)} & \cellcolor{gray!6}{413 (99.5)} & \cellcolor{gray!6}{129 (31.9)} & \cellcolor{gray!6}{245 (71.2)} & \cellcolor{gray!6}{101 (46.3)} & \cellcolor{gray!6}{27 (26.2)} & \cellcolor{gray!6}{1994 (39.2)} & \cellcolor{gray!6}{<0.001}\\
Obesity (morbidly obese) & 106 (9.0) & 22 (2.8) & 46 (8.1) & 50 (9.0) & 15 (2.9) & 33 (8.0) & 22 (5.4) & 56 (16.3) & 28 (12.8) & 13 (12.6) & 391 (7.7) & <0.001\\
\cellcolor{gray!6}{Osteoarthritis} & \cellcolor{gray!6}{301 (25.6)} & \cellcolor{gray!6}{250 (31.5)} & \cellcolor{gray!6}{359 (63.2)} & \cellcolor{gray!6}{107 (19.3)} & \cellcolor{gray!6}{135 (26.1)} & \cellcolor{gray!6}{106 (25.5)} & \cellcolor{gray!6}{118 (29.1)} & \cellcolor{gray!6}{68 (19.8)} & \cellcolor{gray!6}{54 (24.8)} & \cellcolor{gray!6}{18 (17.5)} & \cellcolor{gray!6}{1516 (29.8)} & \cellcolor{gray!6}{<0.001}\\
Osteoporosis & 34 (2.9) & 91 (11.5) & 46 (8.1) & 11 (2.0) & 34 (6.6) & 16 (3.9) & 18 (4.4) & 4 (1.2) & 5 (2.3) & 1 (1.0) & 260 (5.1) & <0.001\\
\cellcolor{gray!6}{Sickle-Cell Anaemia} & \cellcolor{gray!6}{1 (0.1)} & \cellcolor{gray!6}{0 (0.0)} & \cellcolor{gray!6}{0 (0.0)} & \cellcolor{gray!6}{1 (0.2)} & \cellcolor{gray!6}{0 (0.0)} & \cellcolor{gray!6}{1 (0.2)} & \cellcolor{gray!6}{0 (0.0)} & \cellcolor{gray!6}{0 (0.0)} & \cellcolor{gray!6}{0 (0.0)} & \cellcolor{gray!6}{0 (0.0)} & \cellcolor{gray!6}{3 (0.1)} & \cellcolor{gray!6}{0.693}\\
Parkinson's & 23 (2.0) & 17 (2.1) & 11 (1.9) & 4 (0.7) & 8 (1.5) & 10 (2.4) & 5 (1.2) & 4 (1.2) & 0 (0.0) & 0 (0.0) & 82 (1.6) & 0.122\\
\cellcolor{gray!6}{Multiple Sclerosis} & \cellcolor{gray!6}{3 (0.3)} & \cellcolor{gray!6}{0 (0.0)} & \cellcolor{gray!6}{1 (0.2)} & \cellcolor{gray!6}{0 (0.0)} & \cellcolor{gray!6}{0 (0.0)} & \cellcolor{gray!6}{0 (0.0)} & \cellcolor{gray!6}{2 (0.5)} & \cellcolor{gray!6}{1 (0.3)} & \cellcolor{gray!6}{0 (0.0)} & \cellcolor{gray!6}{0 (0.0)} & \cellcolor{gray!6}{7 (0.1)} & \cellcolor{gray!6}{0.409}\\
Epilepsy & 105 (8.9) & 4 (0.5) & 13 (2.3) & 8 (1.4) & 12 (2.3) & 5 (1.2) & 6 (1.5) & 10 (2.9) & 5 (2.3) & 4 (3.9) & 172 (3.4) & <0.001\\
\cellcolor{gray!6}{Dementia} & \cellcolor{gray!6}{85 (7.2)} & \cellcolor{gray!6}{155 (19.5)} & \cellcolor{gray!6}{74 (13.0)} & \cellcolor{gray!6}{37 (6.7)} & \cellcolor{gray!6}{64 (12.4)} & \cellcolor{gray!6}{42 (10.1)} & \cellcolor{gray!6}{31 (7.7)} & \cellcolor{gray!6}{5 (1.5)} & \cellcolor{gray!6}{13 (6.0)} & \cellcolor{gray!6}{4 (3.9)} & \cellcolor{gray!6}{510 (10.0)} & \cellcolor{gray!6}{<0.001}\\
Asthma & 161 (13.7) & 72 (9.1) & 211 (37.1) & 45 (8.1) & 26 (5.0) & 29 (7.0) & 78 (19.3) & 47 (13.7) & 16 (7.3) & 103 (100.0) & 788 (15.5) & <0.001\\
\cellcolor{gray!6}{COPD \textsuperscript{$\S$} } & \cellcolor{gray!6}{227 (19.3)} & \cellcolor{gray!6}{102 (12.9)} & \cellcolor{gray!6}{176 (31.0)} & \cellcolor{gray!6}{63 (11.4)} & \cellcolor{gray!6}{79 (15.3)} & \cellcolor{gray!6}{34 (8.2)} & \cellcolor{gray!6}{97 (24.0)} & \cellcolor{gray!6}{23 (6.7)} & \cellcolor{gray!6}{13 (6.0)} & \cellcolor{gray!6}{25 (24.3)} & \cellcolor{gray!6}{839 (16.5)} & \cellcolor{gray!6}{<0.001}\\
Chronic Pain & 769 (65.4) & 479 (60.4) & 475 (83.6) & 254 (45.8) & 264 (51.1) & 222 (53.5) & 264 (65.2) & 219 (63.7) & 110 (50.5) & 57 (55.3) & 3113 (61.1) & <0.001\\
\cellcolor{gray!6}{Total N (\%)} & \cellcolor{gray!6}{1175 (23.1)} & \cellcolor{gray!6}{793 (15.6)} & \cellcolor{gray!6}{568 (11.2)} & \cellcolor{gray!6}{555 (10.9)} & \cellcolor{gray!6}{517 (10.2)} & \cellcolor{gray!6}{415 (8.1)} & \cellcolor{gray!6}{405 (8.0)} & \cellcolor{gray!6}{344 (6.8)} & \cellcolor{gray!6}{218 (4.3)} & \cellcolor{gray!6}{103 (2.0)} & \cellcolor{gray!6}{5093} & \cellcolor{gray!6}{}\\*
\end{longtable}
\endgroup{}
\end{landscape}

\restoregeometry

%\setcitestyle{aysep={,}}
%\bibliographystyle{apalike}
%\bibliography{/home/marc/Documents/TEX/BIB/LIFE_COURSE.bib,/home/marc/Documents/TEX/BIB/MI.bib,/home/marc/Documents/TEX/BIB/CLUSTERING.bib}  %%% Uncomment this line and comment out the ``thebibliography'' section below to use the external .bib file (using bibtex) .
%\end{document}


\end{document}